# Étude pour l'analyse et l'optimisation du transport des personnes en situation de handicap

Oscar Tellez[a], Laurent Daguet[a], Fabien Lehuédé[b], Thibaud Monteiro[a], Geovanny Osorio Montoya[c], Olivier Péton[b], Samuel Vercraene[a]
[a] Université de Lyon, INSA Lyon, laboratoire DISP EA 4570, Lyon, France.
oscar.tellez@insa-lyon.fr, laurent-daguet@insa-lyon.fr, thibaud.monteiro@insa-lyon.fr, samuel.vercraene@insa-lyon.fr
[b] IMT Atlantique, Laboratoire LS2N, UMR CNRS 6004, Nantes, France.
fabien.lehuede@mines-nantes.fr, olivier.peton@mines-nantes.fr
[c] Ressourcial / Fondation OVE, geovanny.osorio@ressourcial.fr, @fondation-ove.fr

**Résumé.** Depuis 2010, en France, le transport sanitaire est devenu l'une des dix priorités du plan de gestion des risques de l'Assurance maladie du fait de l'augmentation du coût de ces transports. Pour les établissements et services sociaux et médico-sociaux (ESMS), ce coût représente la deuxième dépense après celle du personnel.
Dans ce contexte, le projet de recherche et développement, NOMAd, vise une amélioration globale de la prise en charge du transport quotidien des personnes entre leur domicile et leurs ESMS. À cette fin, nous proposons la mutualisation du transport entre plusieurs ESMS. Cette mutualisation du transport permet de regrouper et d'optimiser les tournées dans une certaine zone géographique. L'enjeu est d'améliorer la performance économique tout en maintenant des objectifs économiques, sociaux et environnementaux.
D'un point de vue scientifique, le problème identifié est nommé le Time-Consistent-Dial-a-Ride Problem et vise à trouver un compromis entre les objectifs de coût de transport et de régularité du service. Étant donné la complexité du problème, nous cherchons, tout d'abord, à résoudre le problème d'une demi-journée en incluant la notion de reconfiguration pendant les tournées. Ensuite, nous considérons la semaine entière. Pour résoudre ces problèmes, nous utilisons la méta-heuristique Large Neighborhood Search et un problème maître basé sur le Set Covering Problem.

**Mots clés :** systèmes d'aide à la décision, maîtrise des flux humains, mutualisation du transport, optimisation, médico-social.

## 1 Introduction

En 2010, le transport sanitaire est devenu l'une des dix priorités du plan de gestion des risques de l'Assurance maladie en France du fait de l'augmentation du coût de ces transports (ANAP, 2013). Pour des établissements médico-sociaux (ESMS), ce coût est en croissance et représente souvent la deuxième dépense après celle du personnel (ANAP, 2013). Ne disposant pas d'une grande expertise en gestion du transport, les ESMS sont souvent menés à adopter des pratiques sous-optimales en termes de coût, de qualité de service ou d'impact environnemental.

À différence du transport sanitaire, le transport médico-social est généralement intégré dans la réalisation du projet de vie des Personnes en Situation de Handicap (PSH) (ANAP, 2013). Le transport n'est plus considéré seulement comme une activité pénible ou sans valeur ajoutée. En effet, il permet d'une part, l'inclusion des PSH, en les emmenant sur leur lieu de travail ou à l'école, et d'autre part, il est un levier important dans l'autonomisation de ces personnes.

Le transport médico-social est en grande partie régulier, car c'est un transport pour lequel les besoins logistiques sont en principe connus à l'avance. Cette connaissance anticipée des besoins peut être utilisée pour améliorer conjointement : la ponctualité et la qualité des prestations de service des bénéficiaires ; la mutualisation des services ; et l'efficience des ressources mobilisées.

La réussite d'un tel projet passe par la mobilisation de tous les acteurs concernés par cette problématique complexe afin de bien prendre en compte les enjeux pour les différentes parties concernées. Nous devons en effet répondre conjointement aux besoins : des personnes en situation de handicap et des aidants familiaux ; des centrales de mobilité et des accompagnateurs ; et des établissements médico-sociaux et des aidants professionnels.

Cette communication s'organise comme suit : En premier lieu nous présentons un positionnement du problème par rapport à la littérature scientifique. Ensuite nous décrivons le projet NOMAd, cadre de ce travail. Après nous présentons notre réflexion sur transport des PSH en deux parties. La section 3 propose une analyse des besoins issue d'une enquête réalisée à l'automne 2017 et une synthèse des caractéristiques spécifiques du transport des PSH. La section 4 expose la démarche utilisée pour organiser ce transport. Cette démarche consiste à proposer des plannings de tournées optimales en coût sur la demi-journée, et à s'en servir pour construire sur une semaine, un planning le plus régulier possible.

## 2 Revue de la littérature

Le problème spécifique du transport à la demande pour les PSH est déjà traité dans la littérature scientifique comme un problème mono-période (Toth & Vigo, 1997 ; Lehuédé et al., 2014). Ce problème prend en compte des éléments spécifiques au transport des PSH comme : la typologie des personnes transportées (fauteuils ou valides) (Parragh, 2011 ; Braekers et al, 2014 ; Masmoudi et al., 2017) et la typologie des véhicules adaptés qui peuvent être configurables (Qu & Bard, 2015), voire reconfigurables (Tellez et al, 2018).

Dès que l'on passe d'un problème mono-période à une situation multi-période, on doit prendre en compte la trajectoire des décisions, notamment la régularité globale du service. Selon la littérature, on peut appréhender la régularité d'une période à une autre de deux façons : la régularité « chauffeur », consistant à minimiser le nombre de chauffeurs différents pour chaque usager (Braekers & Kovacs, 2016) et la régularité « horaire », consistant à réduire la dispersion des horaires de passage d'un même usager. Cette régularité horaire fait l'objet de cette communication.

Le problème de la régularité horaire a déjà été traité dans la littérature scientifique sur un problème de transport assez proche du nôtre, mais plus simple. Appelé "Time-Consistent Vehicle Routing Problem", il recherche donc la régularité horaire et est caractérisé par une flotte de véhicules homogènes, une demande sur un horizon de temps limité et un point d'arrivée unique pour toutes les personnes (Kovacs, 2015 ; Feillet, 2014). Pour notre cas, l'optimisation se fait sur un objectif de régularité défini par le nombre de plages horaires différentes proposées à une personne sur la semaine. Dans notre approche, deux horaires dans la même plage si l'écart est inférieur à 15 minutes. Si une personne a une plage horaire différente chaque jour, il a donc 5 classes. Par conséquent, un planning parfaitement régulier est caractérisé par une classe unique par PSH. Ceci coûte potentiellement très cher. Nous appelons ce problème le time-consistent DARP. Celui-ci n'a, à notre connaissance, pas encore été traité dans la littérature.

## 3 Le projet NOMAd

Le projet « Numérique et Optimisation pour une Mobilité Adaptée » (NOMAd) est un projet de 3 ans soutenu par l'Union Européenne avec le Fond FEDER. Il est déployé sur la région Auvergne Rhône-Alpes. NOMAd vise à développer une application informatique facilitant les échanges entre toutes les parties prenantes du transport adapté : les opérateurs de transport, les personnes transportées (PSH) et les établissements médico-sociaux (voir https://nomad.disp-lab.fr/).

Le but est d'optimiser les plans de transport en intégrant conjointement les points de vue des différents acteurs. Trois éléments sont ciblés :

- La réduction des coûts de transport par la diminution du nombre de kilomètres parcourus et du nombre de véhicules nécessaires ;
- L'amélioration de la qualité de service pour les personnes transportées et les établissements par la réduction des temps de trajet et la régularité des horaires d'un jour sur l'autre ;
- La réduction de l'impact environnemental par la baisse des émissions de $CO_2$ liées au nombre de kilomètres parcourus.

Ce projet est réalisé en partenariat avec deux acteurs du secteur médico-social.

- Ressourcial, qui est un Groupement Social de Moyens (GSM) spécialisé dans les systèmes d'information dédiés au secteur médico-social. Constitué sous forme associative, il est destiné à des ESMS sans but lucratif pour leur permettre le partage d'expériences et la mutualisation d'outils de gestion et d'aide à la décision.
- Le GIHP Service Adapté, qui est spécialisé dans la gestion et la réalisation de transports adaptés aux PSH. Il intervient en particulier dans l'agglomération lyonnaise. Pour donner un ordre de grandeur, cette structure transporte jusqu'à 1500 personnes par demi-journée. Dans cette étude, nous voulons exploiter une particularité du parc de véhicules du GIHP : la possibilité de modifier la capacité du véhicule par rapport aux types de personnes transportées pendant la tournée. Ainsi, il est possible de rabattre des sièges pour permettre d'accueillir plus de personnes en fauteuils roulants pendant une tournée ou inversement, déplier des sièges pour accueillir plus de passagers sans handicap moteur. Cette possibilité est appelée « reconfiguration ».

## 4  Analyse des besoins

### 4.1  Enseignements de l'enquête terrain

En 2017, nous avons réalisé une enquête auprès de 30 directeurs et responsables d'ESMS. Cette enquête donne un aperçu de l'état du transport adapté dans la Région Auvergne Rhône-Alpes.

La Figure 1 montre que le périmètre d'action est caractérisé par une forte dispersion géographique de la demande. Ainsi les ESMS partagent un bassin de recrutement fortement dispersé. Cette caractéristique renforce l'intérêt d'envisager la mutualisation des transports sur plusieurs établissements, mutualisation qui n'est quasiment jamais mise en œuvre dans les pratiques actuelles. Nous pouvons aussi noter que les temps de trajets sont relativement importants. Cette importance est cependant à nuancer, car les cas extrêmes caractérisent des usagers vivants loin de leurs établissements respectifs. Souvent, ces PSH alors sont pensionnaires et leur trajet n'est pas quotidien, mais hebdomadaire. Dans une démarche de mutualisation, ces PSH pensionnaires peuvent être intégrées dans des tournées partagées.

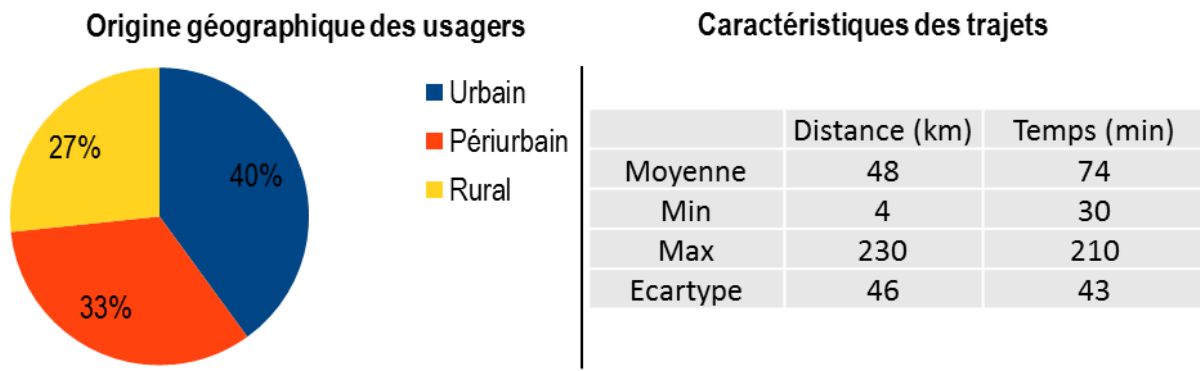

**Figure 1. Origine géographique des usagers**

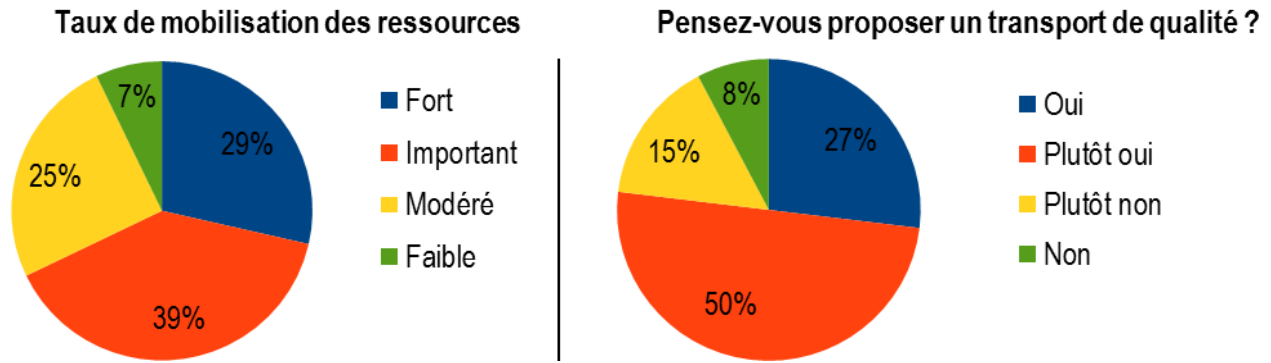

**Figure 2. Implication des établissements**

La Figure 2 montre que les établissements pensent proposer un transport de qualité. Cette perception exprime l'importance pour les ESMS de prendre en compte ce transport, pas seulement comme une nécessaire prestation logistique, mais comme faisant partie intégrante du projet de vie des personnes. Cette importance est aussi traduite par un taux conséquent de mobilisation des personnels. Ceux-ci sont mobilisés pour l'accueil et la préparation au départ des PSH. Elles le sont parfois également pour la réalisation du transport lui-même, si celui-ci n'est pas externalisé. Mais ces ressources sont aussi mobilisées pour la gestion du transport, en recueillant les besoins, en planifiant les tournées et en gérant au quotidien les interactions avec les aidants familiaux. Ces activités sont perçues comme fortement chronophages et sans réelle valeur ajoutée.

Lorsque nous abordons qualitativement la question des principales difficultés rencontrées, les responsables d'établissements pointent les éléments déjà vus ci-dessus, comme la durée et du coût des transports. Ils évoquent plusieurs autres thématiques de préoccupation :

- Premièrement, les responsables d'établissements expriment la nécessité de prendre en compte la spécificité de ce type de transport par son adaptation aux besoins spécifiques des personnes transportées et par la capacité à gérer des troubles de comportements et des risques de crises pendant le transport – avec la difficile question de la contention adaptée pour la sécurité des personnes.
- Deuxièmement, le thème de la communication et de la coordination est souvent relevé, notamment pour la gestion des imprévus, des changements et des retards. Ainsi, la coordination entre les 3 parties prenantes de ce service et la problématique de la bonne transmission des informations est une réelle préoccupation.
- Enfin, les responsables d'établissements évoquent la dimension humaine de ce transport, en insistant sur la mobilisation importante des personnels autour de cette mission et par la formation spécifique des conducteurs eu égard à la qualification et leur stabilité à assurer les tournées.

Concernant les enjeux qui sont à relever, les responsables d'établissements mettent en avant la sécurité et la gestion des risques ; les coûts et l'optimisation de l'organisation ; et l'intégration du transport selon trois perspectives : (1) celle du levier de l'accompagnement à l'autonomie des personnes ; (2) celle du lien entre la famille et l'établissement ; (3) et enfin celle de son articulation avec les autres modes de transports, et notamment l'articulation entre le transport organisé par un opérateur, le transport public et l'utilisation des véhicules personnels.

Au regard des pistes d'actions à proposer, les responsables d'établissements identifient les objectifs suivants :

- développer l'autonomie de la personne ;
- réduire les coûts de fonctionnement et notamment ceux liés à l'entretien des véhicules ;
- réduire les trajets (distance et temps) par la mutualisation ;

- répondre aux besoins d'un territoire défini ;
- faire de la question des transports une facilité plutôt qu'un frein.

Pour conclure ce volet, 16% des établissements interrogés ont intégré des objectifs liés au transport dans la signature de leur contrat d'objectifs et de moyens (CPOM). Dans ces CPOM, cela se traduit par une volonté de maîtriser et de réduire les temps ou les coûts de transport. Parfois est explicitement exprimée une intention de favoriser la mutualisation des circuits avec les autres structures du territoire et le souhait de réorganiser plus souvent les circuits pour gagner en efficience.

Cette enquête terrain nous a permis de définir les trois objectifs prioritaires du projet NOMAd. Il s'agit de la maîtrise des coûts, de la qualité de service et de la capacité à communiquer et se coordonner dans des actions de mutualisation.

## 4.2 Un problème à dimensions multiples

La gestion et l'optimisation du transport de personnes en situation de handicap sont complexes. Il nécessite de prendre en compte un ensemble de variables et de spécificités, lié à la fois aux personnes, aux véhicules et au schéma d'organisation.

**Personnes bénéficiaires.** Les personnes bénéficiaires sont de plusieurs types. Elles peuvent être en fauteuils ou peuvent marcher. Elles ont un certain degré d'autonomie ; mais certaines doivent être accompagnées par un tiers.
Dans un objectif de qualité de service, la dimension horaire est aussi importante. Les acteurs présentent des contraintes de disponibilités qui doivent être respectées. Ces contraintes sont exprimées sous forme de disponibilités horaires au départ et à l'arrivée. Une durée maximale de transport pour chaque personne est prise en compte. Enfin, la régularité horaire est aussi à intégrer lors de l'élaboration d'un plan de transport. En effet, en dépit de la variabilité de la demande sur la semaine, on préférera donner aux personnes présentes plusieurs jours un unique horaire de passage du véhicule le matin et le soir.

**Véhicules.** Les véhicules utilisés pour le transport des personnes à mobilité réduite sont aménagés spécifiquement. Nous considérons une flotte de véhicules hétérogènes. Chacun est caractérisé par une capacité d'accueil de différents types de personnes et par une capacité à pouvoir être configuré pour un type de capacité spécifique avant le départ du véhicule du dépôt ou à pouvoir avoir cette capacité reconfigurée en cours de journée. Cette reconfiguration est, par exemple, réalisée avec des sièges relevables, permettant notamment de transporter alternativement des personnes en fauteuil roulant ou des personnes pouvant marcher.

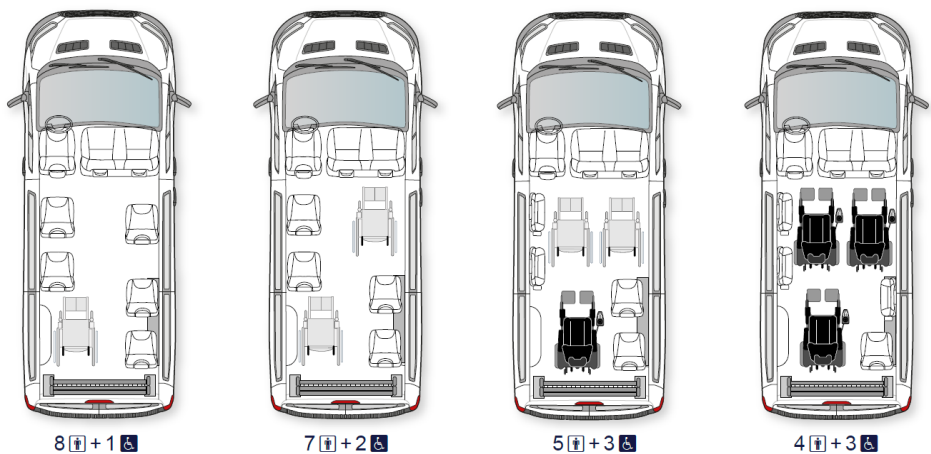

Figure 3. Exemples d'un véhicule pouvant être reconfiguré (https://www.handynamic.fr/)

La Figure 3 montre quatre types de capacités différentes pour un véhicule reconfigurable. Les capacités sont caractérisées ici par des places classiques, des places pour fauteuil simple et des places pour fauteuil électrique.

**Co organisation.** Les différentes parties prenantes sont caractérisées par des objectifs différents, voire antagonistes. En effet, une organisation d'un plan de transport basée uniquement sur un critère financier aura tendance à réduire le nombre de véhicules mobilisés et à proposer des trajets plutôt longs. Ce qui d'un point de vue de la qualité de service n'est pas optimal. À l'inverse, une organisation d'un plan de transport cherchant uniquement à proposer la meilleure qualité de service aura tendance à multiplier le nombre de véhicules utilisés et à proposer des transports en trajet direct pour chaque personne. Aussi, il semble nécessaire de trouver une solution de compromis. Une approche multicritère est donc requise pour prendre à la fois des critères de réduction des coûts et d'impact environnemental et des critères de qualité de service.

Du point de vue de l'organisation logistique, une approche que nous souhaitons mettre en place est celle de la mutualisation. Les tournées sont aujourd'hui planifiées établissement par établissement. Il serait plus efficace de permettre le transport dans un même véhicule de personnes d'établissements différents. Néanmoins cette mutualisation augmente significativement la difficulté de la planification manuelle.

## 5 Construction du planning hebdomadaire

Dans le cadre de ce travail, nous faisons l'hypothèse que le programme de transport est construit à partir d'un canevas constitué de semaines types. Cette section présente la démarche de construction d'une semaine type. Définir un planning hebdomadaire est une tâche complexe. Le problème a donc été décomposé en deux sous-problèmes. Dans un premier temps, nous nous sommes intéressés à la planification d'une brique élémentaire, à savoir une demi-journée. Dans un second temps, nous nous sommes penchés sur la construction du planning hebdomadaire, à partir des briques élémentaires qui le compose. La Figure 4 illustre la démarche. Le rectangle M1 correspond à une des briques élémentaires relatives aux décisions de demi-journée. Le rectangle en pointillé précise le périmètre du planning hebdomadaire.

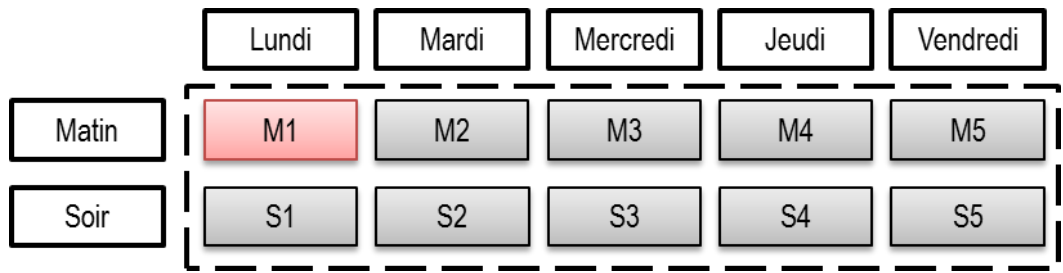

**Figure 4. Structuration d'une semaine type**

### 5.1 Le planning de la demi-journée

Définir le planning de la demi-journée, brique élémentaire du planning hebdomadaire, consiste à résoudre une extension d'un problème de logistique connu sous le terme de Dial-a-Ride Problem (DARP). Le DARP est caractérisé par un point de départ et d'arrivée différents pour chaque personne transportée, des horaires de ramassage et de dépose (fenêtre de temps) et une durée maximale de transport. Des contraintes liées aux places des véhicules sont aussi intégrées.

L'optimisation se fait sur un objectif de coût qui intègre l'amortissement des véhicules, des coûts horaires (ex : coût des conducteurs-accompagnateurs), et des coûts kilométriques. Les respects des horaires de passage et des limites de temps de séjour dans le véhicule sont exprimés sous forme de contraintes.

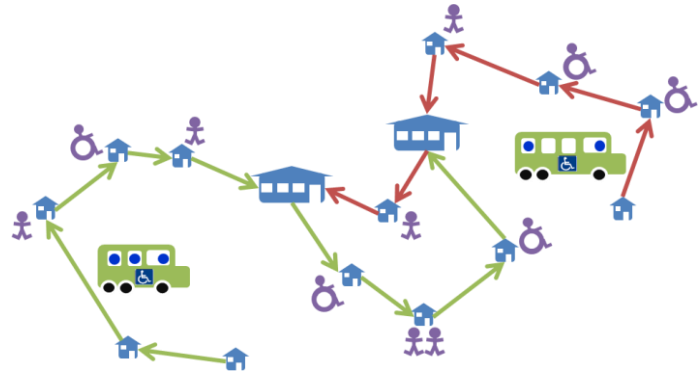

**Figure 5. Illustration de tournées**

La Figure 5 illustre deux tournées du matin consistant à transporter des PSH vers leurs établissements. Ces deux tournées déposent les personnes dans deux établissements, montrant ici l'intérêt de la mutualisation.

## 5.2 Régularité sur une semaine

Une fois résolu le problème du transport d'une demi-journée, nous procédons à une résolution hebdomadaire. En effet, toutes les personnes ne demandent pas un transport chaque jour vers leur établissement. La simple juxtaposition des plans de transport de toutes les demi-journées donne un planning très efficace en coût. Mais, cela signifie que des personnes ont potentiellement un horaire de prise en charge différent chaque jour. Cette situation est, a minima, inconfortable pour les personnes, voire pleinement, inacceptables pour certains d'entre eux. En effet, certaines PSH atteintes d'un handicap mental ne tolèrent que difficilement un changement d'habitude. Ainsi, dans un objectif de qualité de service et de maîtrise des variations horaires, un planning hebdomadaire doit viser à la fois une maîtrise des coûts et une régularité mesurée sur l'ensemble de la semaine.

La Figure 6 illustre l'approche globale de résolution du time-consistent DARP. Nous commençons par résoudre le sous-problème de chaque demi-journée décrit en la section 4.1. L'union de ces solutions constitue le point de départ de la construction du planning hebdomadaire. Agrégeant les solutions économiquement proches de l'optimum, celui-ci est performant en coût, mais pas en régularité. Pendant cette première étape, nous collectons les tournées obtenues dans les différentes itérations pour constituer une réserve de tournées utiles pour l'amélioration de la régularité. Ensuite, avec l'aide d'un moteur d'optimisation nous résolvons, de façon itérative, le problème du time-consistent DARP pour améliorer la régularité en utilisant chaque fois un sous-ensemble de tournées de la réserve. Ce processus est répété jusqu'à l'obtention de la régularité désirée.

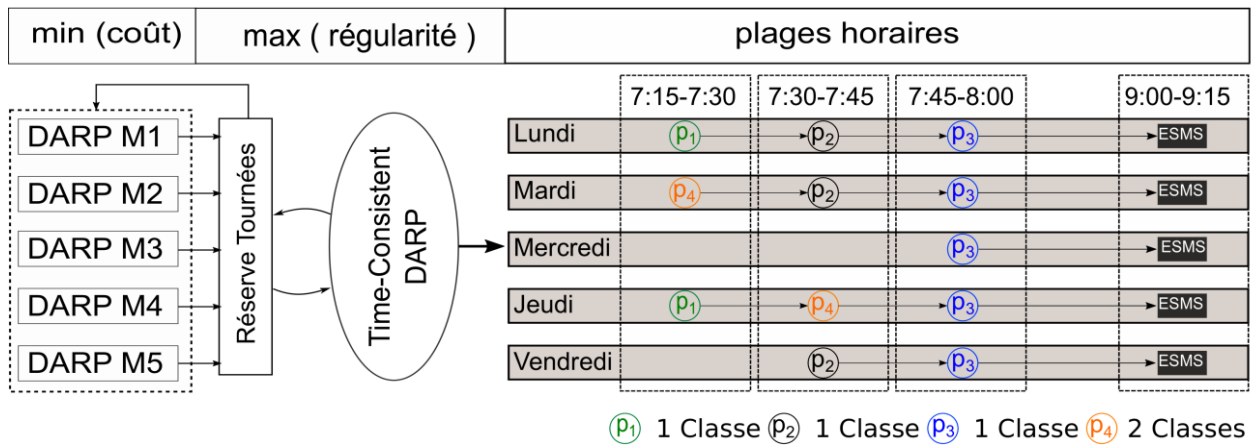

**Figure 6. Approche globale du processus de décision**

## 6   Bilan

La gestion et l'optimisation du transport des PSH sont complexes. Il nécessite de prendre en compte un ensemble de variables et de spécificités, liées à la fois aux personnes, aux véhicules et aux schémas d'organisation. Dans les ESMS, ce processus demande très souvent une mobilisation importante des ressources humaines pour assurer un transport de qualité. Sachant que ce transport est caractérisé par une forte dispersion géographique des demandes, il y a un vrai intérêt économique à optimiser et mutualiser le service. Ceci doit être fait sans dégrader la qualité de service. À cette fin, nous proposons une approche globale de résolution pour la planification du transport à la fois efficace en coût et régulier en temps de service pour chacune des personnes transportées.

## 7   Remerciement

## 8   Bibliographie